\documentclass[preprint,aps,12pt,preprintnumbers,eqsecnum,nofootinbib]{revtex4}
\usepackage{xcolor}
\usepackage{graphicx}
\usepackage{subfigure}
\usepackage{epstopdf} 
\usepackage{braket}
\usepackage{amssymb,amsmath,amsfonts,comment,latexsym,graphicx,epstopdf,bbold,float}
\usepackage{color}
\usepackage{fancyvrb}
\usepackage{chngcntr}
\usepackage{etoolbox}
\usepackage{bbm,epsfig}
\usepackage{array}
\usepackage{hepunits}
\usepackage{slashed}

\newcommand{\be}{\begin{equation}}
\newcommand{\ee}{\end{equation}}
\newcommand{\ba}{\begin{eqnarray}}
\newcommand{\ea}{\end{eqnarray}}

\newcommand{\grts}{\raise.3ex\hbox{$>$\kern-.75em\lower1ex\hbox{$\sim$}}}
\newcommand{\lets}{\raise.3ex\hbox{$<$\kern-.75em\lower1ex\hbox{$\sim$}}}

{\catcode`\|=\active\gdef\Braket#1{\left<\mathcode`\|"8000\let|\bravert {#1}\right>}}

\def\bravert{\egroup\,\vrule\,\bgroup}

\usepackage{subfigure}

\usepackage{color}
\usepackage{amssymb,amsmath,amsfonts}
\usepackage{epstopdf} 
\usepackage{braket}

\begin{document}
%
%
\title{\vspace*{0.5in} 
Aspects and applications of nonlocal Lorentz-violation
\vskip 0.1in}
\author{Christopher D. Carone}\email[]{cdcaro@wm.edu}
\affiliation{High Energy Theory Group, Department of Physics,
William \& Mary, Williamsburg, VA 23187-8795}

%
%
\date{August 24, 2020}
\begin{abstract}
We consider simple scalar theories with quadratic terms that are nonlocal and Lorentz violating.  Unlike similar Lorentz-invariant 
nonlocal theories that we have considered previously, the theories studied here are both ghost-free and unitary as formulated in Minkowski 
space.  We explore the possibility that the scale of nonlocality could be low in a dark sector, where the stringent constraints on the violation
of Lorentz invariance may be accommodated via the weak coupling to the standard model.   We point out that long-range forces may originate 
from such a sector and be distinguishable from more conventional beyond-the-standard model possibilities.  We present a model in which a 
nonlocal, Lorentz-violating dark sector communicates with the standard model via a sector of heavy vector-like fermions, a concrete framework 
in which phenomenological constraints and signals can be investigated.
\end{abstract}

\maketitle

\section{Introduction} \label{sec:intro}

Quantum field theories involving nonlocal interactions are interesting for a variety of reasons~\cite{manyapps,Carone:2016eyp}.  For example, 
in Ref.~\cite{Carone:2016eyp}, we studied a nonlocal, Lorentz-invariant theory of $N$ real scalar fields of mass $m$ with O($N$) symmetry 
\begin{equation}
{\cal L} =
- \frac{1}{2} \phi^a \, \hat{F}(\Box)^{-1}  (\Box+m^2) \, \phi^a -\frac{1}{8} \lambda_0 (\phi^a\phi^a)^2 \,\,\, .
\label{eq:LImodel}
\end{equation}
Here $a=1\ldots N$, $\lambda_0$ is the dimensionless quartic coupling, and
\begin{equation}
\hat{F}(\Box) = \exp(-\eta\Box^n) \,\, ,
\label{eq:LIF}
\end{equation}
where our metric signature is $(+,-,-,-)$. The parameter $\eta$ determines the amount of nonlocality, with the local theory 
corresponding to $\eta=0$ and $\hat{F}=1$.   One reason that this theory is of interest is that the propagator 
\begin{equation}
\tilde{D}_F(p) = \frac{i \, \hat{F}(-p^2)}{p^2 -m^2 + i \epsilon} 
\label{eq:LIprop}
\end{equation}
leads to more convergent amplitudes than in the local theory with $\hat{F}=1$; for $n$ even, this is true whether the theory is formulated initially in Minkowski or 
Euclidean space.  While better convergence properties can also be obtained in local theories with higher-derivative quadratic terms, like the Lee-Wick Standard 
Model~\cite{Grinstein:2007mp}, such theories unavoidably come with ghosts; special prescriptions must then be invoked in computing $S$-matrix elements to 
maintain the unitarity of the theory~\cite{Cutkosky:1969fq}.  These extra theoretical ingredients are arguably unappealing, but can be avoided in the nonlocal 
theory above if $\hat{F}$ is chosen to be an entire function, as in Eq.~(\ref{eq:LIF}), so that no new poles appear in the propagator, Eq.~(\ref{eq:LIprop}).   If such 
an approach could be generalized convincingly to gauge theories, one hope is that a nonlocal generalization of the standard model could be used to address the 
hierarchy problem, without implying new, TeV-scale particles that have yet to be seen at the Large Hadron Collider (LHC).

Complications related to unitarity, however, also arise in the ghost-free theory defined by Eqs.~(\ref{eq:LImodel}) and
(\ref{eq:LIF}).  In Ref.~\cite{Carone:2016eyp}, two-into-two scattering was considered to all orders in the quartic coupling in the large $N$ limit, 
and it was shown that the theory with $n=2$ was not unitary if it is defined in Minkowski space. The problem originates from the form 
of $\hat{F}(-p^2)$, which blows up within certain Stokes wedges in the complex $p^0$ plane; this leads to new 
contributions to the imaginary part of the forward scattering amplitude (coming from the contour at infinity) that 
would not be present after a Wick rotation in the $\hat{F}=1$ version of the theory.  This problem seems to be generic for 
other choices of $\hat{F}$ that are entire functions. On the other hand, one can define the nonlocal theory initially in Euclidean 
space and analytically continue scattering amplitudes to Minkowski space at the very end.  In this case, the resulting theory was 
shown to satisfy the optical theorem~\cite{Carone:2016eyp}.  However, this formulation may seem as unappealing to some as the 
special prescriptions employed to render Lee-Wick theories consistent.

In the present work, we avoid these complications by considering similar nonlocal theories that are not Lorentz-invariant.\footnote{For a very different approach to 
nonlocal Lorentz-violation, See Ref.~\cite{Altschul:2004wr}.}  We explore a simple 
modification of our previous choice for $\hat{F}$, in which d'Alembertian operator is replaced by the Laplacian:
\begin{equation}
\hat{F}(\nabla) = \exp(\eta\,\nabla^2) \,\, .
\label{eq:LVF}
\end{equation}
The theory defined with this operator is Lorentz violating; there is a preferred frame in which the Lagrangian is invariant  
under spatial rotations.  One possible choice is to assume that this is the frame in which the cosmic microwave background is  
isotropic~\cite{Coleman:1998ti}, though others are imaginable\footnote{Other assumptions for a preferred frame that have
appeared in the literature include ones at rest with respect to our Galaxy and that locally co-move with the rotation of the 
Galaxy~\cite{Shao:2012eg}, or co-move with the Barycentric Celestial Reference System~\cite{Soffel:2008zz}.}; motion relative to 
the preferred frame, which introduces a preferred direction, can be separately bounded.  In any case, the absence of time derivatives 
in Eq.~(\ref{eq:LVF}) eliminates the problem with unitarity encountered in the Minkowski-space formulation of the Lorentz-invariant theory 
defined by Eqs.~(\ref{eq:LImodel}) and (\ref{eq:LIF}); it also assures that there are no ghosts, as the inverse propagator involves no higher 
powers of $p^0$.   

If the nonlocality represented by Eq.~(\ref{eq:LVF}) is relevant in nature, one would expect that modification of gauge-invariant quantities 
in the standard model (for example, $-\frac{1}{4}\, B_{\mu\nu} \,\hat{F}(\nabla)^{-1} \,B^{\mu\nu}$, where $B_{\mu\nu}$ is the hypercharge field 
strength tensor) would lead to significant lower bounds on the nonlocality scale $\eta^{-1/2}$ due to the stringent experimental constraints on the violation 
of Lorentz invariance~\cite{Tasson:2014dfa,Kostelecky:2008ts}.  As a consequence, solving the hierarchy problem would not be a motivation for studying 
such theories.  However, there are other motivations for why nonlocality may be relevant in nature (for example, in smoothing out gravitational 
singularities~\cite{smeared}) and the scale of the nonlocality need not be the same for every particle.   One interesting possibility is the application of 
Lorentz-violating nonlocality to gravitation, a nonlocal generalization of the Horava-Lifshitz idea~\cite{Horava:2009uw}, motivated by the desire to obtain a 
renormalizable quantum theory of gravity.  Another interesting possibility is that the nonlocality scale associated with a dark sector may be much lower than the Planck scale,  with the bounds on Lorentz violation accommodated by a very weak coupling of the dark sector to the standard 
model.\footnote{We are not imagining that the nonlocality would be unique to the dark sector, only that its effects may be more accessible there since 
this is the sector of the theory where the constraints are weakest.}  As a first step in model building, we study a nonlocal, Lorentz-violating dark sector 
later in this paper, and defer the consideration of gravity to future work.   With the extremely small couplings required, an interesting 
phenomenological possibility is that dark sector particles may mediate long-range forces.  In this case, effects might be discerned relative to the  
effects of gravity, and lead to corrections to the gravitational potential that differ qualitatively from other possibilities that have been considered previously~\cite{Fischbach:1996eq}.

Our paper is organized as follows.  In the next section, we revisit the analysis of unitarity that was discussed in Ref.~\cite{Carone:2016eyp}, and show how it is 
modified, in a favorable way, for the choice of higher-derivative terms given by Eq.~(\ref{eq:LVF}).  In Sec.~\ref{sec:potential}, we consider the nonrelativistic 
potential for a single scalar field with the same nonlocal Lagrangian, and show how it differs qualitatively from that of the corresponding local theory.  While this 
calculation is based on the assumption that the scalar has  generic Yukawa couplings to generic fermions, in Sec.~\ref{sec:darksector} we present a scenario 
that provides an origin for the weak couplings to standard model fermion fields, by connecting the dark and visible sectors via a renormalizable and gauge
invariant ``portal" sector of heavy, vector-like fermions.  With an explicit scenario defined, we consider the implications of searches for long range forces and for 
the violation of Lorentz invariance on the mass scale and couplings associated with the vector-like sector, assuming that the scale of nonlocality is comparable to
the mass scale of the particle mediating the long-range force.   In the final section, we summarize our conclusions.

\section{Unitarity in a toy model} \label{sec:unitarity}
In Ref.~\cite{Carone:2016eyp}, unitarity in a two-into-two scattering process was considered in the model defined by Eqs.~(\ref{eq:LImodel}) and (\ref{eq:LIF}).  
The calculation was done in the large $N$ limit, where the result at leading order in $1/N$ could be conveniently re-summed, to all orders in perturbation theory.  
In this section, we revisit that calculation and show how it is modified with the form of $\hat{F}$ given in Eq.~(\ref{eq:LVF}).

We consider the two-into-two scattering amplitude ${\cal M} (ab \rightarrow cd)$, for the diagrammatically simplest case in which $a=b\neq c=d$, where field labels $a$ through $d$ 
range from $1 \ldots N$.   The scattering amplitude is given by
\begin{equation}
{\cal M} = - \frac{\lambda}{N}  \frac{e^{-\frac{1}{2} \eta (|\vec{k}_1|^2 + |\vec{k}_2|^2 + |\vec{k}'_1|^2 + |\vec{k}'_2|^2)}}{1 + \lambda\, \Sigma(s)} \, \delta_{ab} \delta_{cd} \,\,\, ,
\label{eq:scatamp}
\end{equation}
where $\lambda_0 \equiv \lambda/N$ to make the large-$N$ scaling of the amplitude explicit.  We indicate the momenta of the incoming scalar bosons by $k$ and the outgoing ones by $k'$.  This amplitude resums all orders 
in $\lambda$ at leading order in $1/N$.  Eq.~(\ref{eq:scatamp}) should be compared to Eq.~(2.8) in Ref.~\cite{Carone:2016eyp}; the same sign convention for self-energy function $\Sigma$ is used.  
The differing numerator corresponds to the differing wave function renormalization factors on each of the four external lines.  The function $\Sigma$ in the present case is given by
\begin{equation}
\Sigma = -\frac{i}{2} \int \frac{d^4 k}{(2 \pi)^4}
\frac{\exp\{-\eta\, (\vec{k}+\vec{p}/2)^2\}\exp\{-\eta\, (\vec{k}-\vec{p}/2)^2\}}
{[(k-p/2)^2 -m^2 + i \epsilon] [(k+p/2)^2-m^2 + i \epsilon]} \,\,\, ,
\label{eq:sigma}
\end{equation}
where $p \equiv k + k'$.

For a single scalar field, the optical theorem relates the total scattering cross section to the imaginary part of the forward scattering amplitude, where $k'_i=k_i$.  In the 
present case where there are $N$ fields with $a=b\neq c=d$, the optical theorem requires
\begin{eqnarray}
&&2\, {\rm Im } \, {\cal M}( k_1,a; k_2, a \rightarrow k_1, c ; k_2, c)   \nonumber \\
&&=\frac{1}{2} \sum_f  \int \frac{d^3 q_1}{(2 \pi)^3}
\frac{d^3 q_2}{(2 \pi)^3} \frac{1}{2 E_1} \frac{1}{2 E_2} (2 \pi)^4 \delta^{(4)}(q_1+q_2-k_1-k_2) \nonumber \\
&& \times \, {\cal M}(k_1,a; k_2,a \rightarrow q_1,f;q_2, f) \nonumber \\
&& \times \, {\cal M}^*(k_1,c; k_2,c \rightarrow q_1,f;q_2, f) \,\,\, .
\label{eq:ot}
\end{eqnarray}
We now prove the equality of the left- and right-hand sides of Eq.~(\ref{eq:ot}).   The imaginary part of the forward scattering amplitude is proportional to the imaginary part of $\Sigma$:
\begin{equation}
2\, {\rm Im } \, {\cal M}( k_1; k_2 \rightarrow k_1; k_2) = \frac{2 \lambda^2}{N} {\rm Im}\, \Sigma
\frac{\exp\{-\eta (|\vec{k}_1|^2 +|\vec{k}_2|^2)\}}{|1+\lambda \Sigma|^2} \,\,\,.
\label{eq:impartM}
\end{equation}
We evaluate the $k^0$ integral in Eq.~(\ref{eq:sigma}) by closing a contour in the lower half of the complex
$k^0$ plane; from the $i \epsilon$ prescription, this encloses poles at $k^0=E_{\vec{k}-\vec{p}/2}+p^0/2$ and $E_{\vec{k}+\vec{p}/2}-p^0/2$, where $E^2_{\vec{q}} \equiv \vec{q\,}^2+m^2$.  In textbook treatments of the optical theorem, one generally works in a frame where $\vec{p}=0$.  In the present case, such a boost away from the preferred frame would also change the form of the Lagrangian (which is not Lorentz invariant), reintroducing $k^0$ dependence into the numerator of Eq.~(\ref{eq:sigma}); this would not be desirable for the reasons related to Wick rotation described earlier.  Hence we work with the Eq.~(\ref{eq:sigma}) in the preferred frame and will comment later on how one could have approached the problem starting in a different frame. Using the residue theorem, one obtains
\begin{eqnarray}
\Sigma &=& - \frac{1}{2} \int \frac{d^3 k}{(2 \pi)^3} \,\,
\frac{\exp\{-\eta\, (\vec{k}+\vec{p}/2)^2\}\exp\{-\eta\, (\vec{k}-\vec{p}/2)^2\}}
{E_{\vec{k}-\vec{p}/2}-E_{\vec{k}+\vec{p}/2}+p^0}   \nonumber \\
&&\left[\frac{1}{2 E_{\vec{k}-\vec{p}/2}(E_{\vec{k}-\vec{p}/2}+E_{\vec{k}+\vec{p}/2}+p^0)} -
\frac{1}{2 E_{\vec{k}+\vec{p}/2}(E_{\vec{k}-\vec{p}/2}+E_{\vec{k}+\vec{p}/2}-p^0)} \right] \,\,\,.
\end{eqnarray}
The imaginary part of $\Sigma$ is related to the branch cut singularity originating from the second term in brackets\footnote{Note that the first term in brackets and the integrand prefactor have no singularities.  In the latter case, this can be seen by noting that as a function of $|\vec{k}|$, the quantity $E_{\vec{k}+\vec{p}/2} - E_{\vec{k}-\vec{p}/2}$ is no larger than $|\vec{p}|$, which is always less that $p^0$ when expressed in terms of the on-shell external momenta,  $p =k_1 + k_2$.}. The discontinuity from crossing this singularity in the complex $p^0$ plane is related to the imaginary part by 
${\rm Disc} \, \Sigma = 2 i {\rm Im}\, \Sigma$.   Moreover, we may use the identity
\begin{equation}
{\rm Disc} \, \frac{1}{p^0 -(E_{\vec{k}-\vec{p}/2}+E_{\vec{k}+\vec{p}/2})} = -2 i \pi \,
\delta(p^0 -E_{\vec{k}-\vec{p}/2}-E_{\vec{k}+\vec{p}/2})
\end{equation}
This allows us to write
\begin{equation}
{\rm Im}\, \Sigma = - \frac{\pi}{2} \int \frac{d^3k}{(2 \pi)^3}  \frac{\exp\{-\eta\, (\vec{k}+\vec{p}/2)^2\}\exp\{-\eta\, (\vec{k}-\vec{p}/2)^2\}}
{(2 E_{\vec{k}+\vec{p}/2})(E_{\vec{k}+\vec{p}/2}-E_{\vec{k}-\vec{p}/2}-p^0)} \delta(p^0 -E_{\vec{k}-\vec{p}/2}-E_{\vec{k}+\vec{p}/2})  \,\,\, ,
\end{equation}
or in the more suggestive form
\begin{equation}
{\rm Im}\, \Sigma =  \frac{1}{4} \int \frac{d^3k}{(2 \pi)^3} \frac{\exp\{-\eta\, (\vec{k}+\vec{p}/2)^2\}\exp\{-\eta\, (\vec{k}-\vec{p}/2)^2\}}
{(2 E_{\vec{k}+\vec{p}/2}) (2 E_{\vec{k}-\vec{p}/2})} (2 \pi) \delta(p^0 -E_{\vec{k}-\vec{p}/2}-E_{\vec{k}+\vec{p}/2}) \,\,\,.
\end{equation}
It follows from Eq.~(\ref{eq:impartM}) that the left-hand-side (LHS) of the optical theorem can be written
\begin{eqnarray}
&&LHS = 2\, {\rm Im } \, {\cal M}( k_1; k_2 \rightarrow k_1; k_2) = \frac{\lambda^2}{2 \, N} 
\frac{1}{|1+\lambda \Sigma|^2} e^{-\eta (|\vec{k}_1|^2 +|\vec{k}_2|^2)} \nonumber \\
&&
\int \frac{d^3k}{(2 \pi)^3} \frac{1}{2 E_{\vec{k}+\vec{p}/2}}\frac{1}{2 E_{\vec{k}-\vec{p}/2}}
(2 \pi) \delta(p^0 -E_{\vec{k}-\vec{p}/2}-E_{\vec{k}+\vec{p}/2}) 
e^{-\eta\, (\vec{k}+\vec{p}/2)^2}e^{-\eta\, (\vec{k}-\vec{p}/2)^2}
\,\,\,.
\label{eq:lhs}
\end{eqnarray}
To evaluate the right-hand-side (RHS) of the optical theorem, Eq.~(\ref{eq:ot}), we write $p\equiv k_1+k_2$ and note that $\Sigma$ is a function of $p$ and can be pulled outside the integrals.  Hence,
\begin{equation}
RHS = \frac{\lambda^2}{2 \, N}\frac{1}{|1+\lambda \Sigma|^2} e^{-\eta (|\vec{k}_1|^2 +|\vec{k}_2|^2)}
\int \frac{d^3 q_1}{(2 \pi)^3} \frac{d^3 q_2}{(2 \pi)^3} \frac{1}{2 E_1} \frac{1}{2 E_2} (2 \pi)^4 
\delta^{(4)}(q_1+q_2-p) e^{-\eta(|\vec{q}_1|^2 +|\vec{q}_2|^2)} \,.
\label{eq:rhs}
\end{equation}
Notice that the prefactors multiplying the integrals in Eq.~(\ref{eq:lhs}) and (\ref{eq:rhs}) coincide.  Hence, we focus on the integral in Eq.~(\ref{eq:rhs}).  First, we may do the $d^3 q_2$ integral using the three-dimensional delta-function.  Since the $q_0^i$ are on shell, this makes the remaining delta-function a function of 
$q_1^0 \equiv E_1= E_{\vec{q}_1}$ and $q_2^0 \equiv E_2=E_{\vec{p}-\vec{q}_1}$.  Next, we shift the remaining
integration variables, $\vec{q}_1 \rightarrow \vec{q}_1+\vec{p}/2$, so that the RHS integral becomes
\begin{equation}
\int \frac{d^3 q_1}{(2 \pi)^3} \frac{1}{2 E_{\vec{q}_1+\vec{p}/2}} \frac{1}{2 E_{\vec{q}_1-\vec{p}/2}} (2 \pi)
\delta(E_{\vec{q}_1+\vec{p}/2}+E_{\vec{q}_1-\vec{p}/2} -p^0) e^{-\eta\, |\vec{q}_1+\vec{p}/2|^2}
e^{-\eta\, |\vec{q}_1-\vec{p}/2|^2} \,.
\end{equation}
With the relabeling $q_1 \rightarrow k$, both the prefactors and integrals in LHS and RHS agree, showing that 
the optical theorem is satisfied.

It is worth noting that agreement between the LHS and RHS of the optical theorem would not have been spoiled had 
we worked in a frame where the $\nabla^2$ of Eq.~(\ref{eq:LVF}) were replaced by the more general form 
$N_{\mu\nu} \partial^\mu\partial^\nu$, with $N_{\mu\nu} = \delta_{ij} \, {\Lambda^i}_\mu \, {\Lambda^j}_\nu$,  where 
$\Lambda$ is an appropriate Lorentz transformation matrix that connects the preferred frame to a given one.   While the 
form of the Lagrangian in the non-preferred frame will change the exponential factors that
appear at the starting points of the previous LHS and RHS derivations, one would, in the very next step, use the 
Lorentz invariance of the remaining factors in the integrands (and integration measures) to change variables so that 
the exponential factors again depend only on $\vec{k}$.  If one were to express the external momenta in terms of their 
values in the preferred frame, then the calculation would be identical to the one just presented.  

\section{Nonrelativistic potential for long-range forces} \label{sec:potential}
The nonlocality defined by Eq.~(\ref{eq:LVF}) violates Lorentz invariance, a possibility that is tightly bounded by 
experiment~\cite{Tasson:2014dfa}.  As we indicated earlier, such a nonlocal modification of the standard model would lead to a high nonlocality 
scale; however, the nonlocality in a dark sector that is adequately sequestered from the standard model could come at a lower scale, due to the small
coupling between the two sectors.  We explain in Sec.~\ref{sec:darksector} how we can induce such small couplings between a Lorentz-violating, nonlocal dark sector and standard model fermions. In this section, we will assume such an effective coupling $g$ exists, in the form of a Yukawa interaction between a single dark-sector scalar field $\phi$ and a generic fermion $\psi$: 
\begin{equation}
{\cal L}_{int} = - g \, \overline{\psi} \, \psi \, \phi \,\,\, .
\end{equation}
We will show in Sec.~\ref{sec:darksector} that the bounds on Lorentz violation force $g$ to be extremely small, far too small to look for effects in any existing 
collider experiments.  However, such small couplings, like that of gravity, can have observable effects when macroscopic quantities of matter are involved, and the 
effect of the scalar is suitably long ranged.   Hence, in this section, we consider such a nonlocal long-ranged force.   While the exponential factor in 
the $\phi$ Lagrangian regulates potential for the long-range force in the ultraviolet, that ultraviolet scale does not necessarily have to be very high if the 
coupling to standard model particles is weak.  This can lead to changes in the shape of the potential at length scales where differences might be 
discernible in comparison to more conventional possibilities.

\begin{figure}[t]
 \begin{center}
   \includegraphics[width=.5\textwidth]{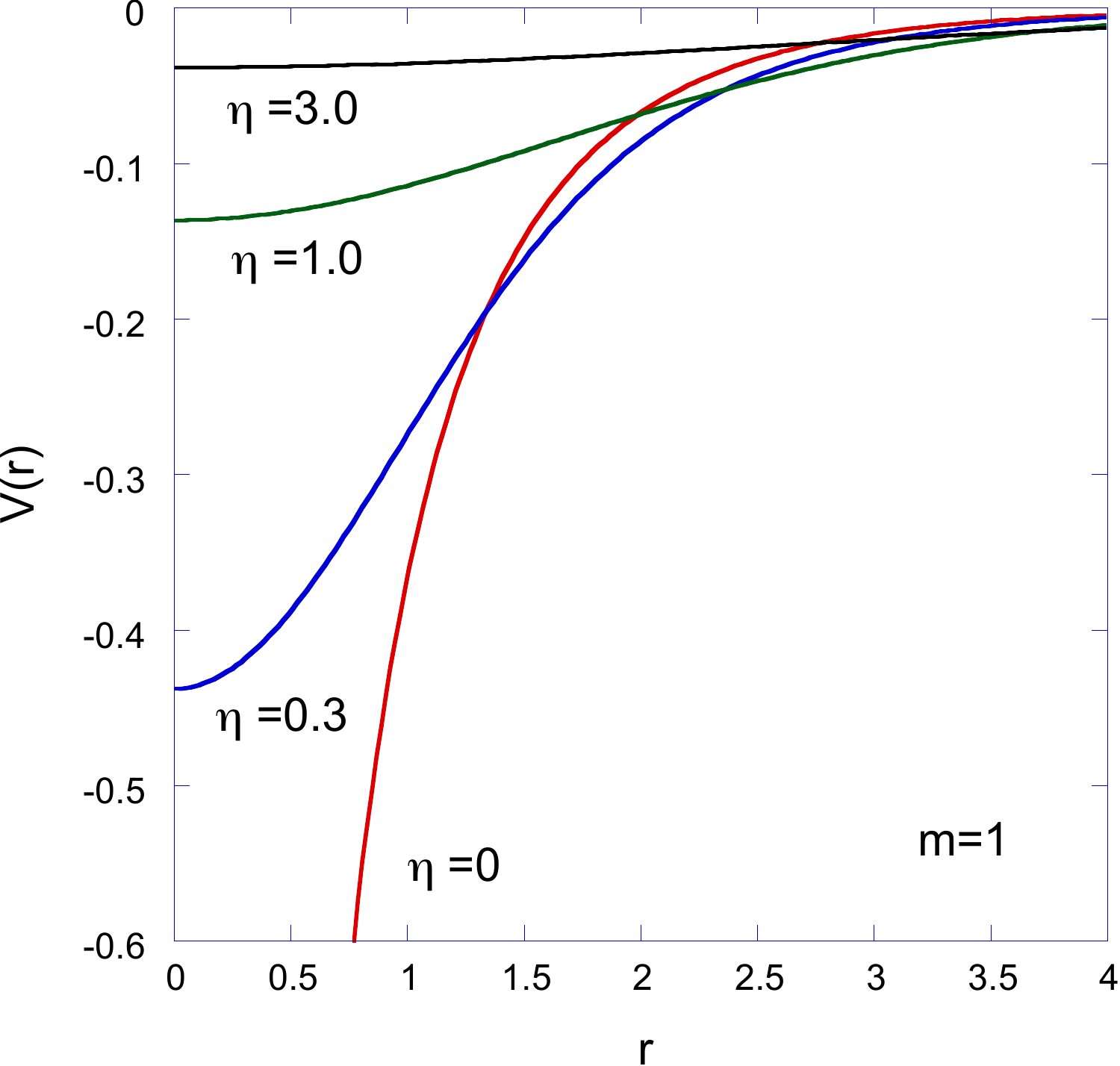}
   \caption{The form of the nonrelativistic potential for various $\eta$, with $m$ set equal to $1$.}
        \label{fig:one}
  \end{center}
\end{figure}

The nonrelativistic potential can be computed in a quantum field theory via an expression proportional to the Fourier transform of the propagator of the force-carrying particle in the nonrelativistic limit.  For example, in ordinary Yukawa theory
\begin{equation}
V(\vec{x}) = \int \frac{d^3 q}{(2 \pi)^3} \frac{ -g^2}{|\vec{q}|^2 + m^2} e^{i \, \vec{q}\cdot \vec{x}}  
=-\frac{g^2}{4 \pi^2\, i r} \int_{-\infty}^\infty d |\vec{q}|  |\vec{q}| \frac{1}{|\vec{q}|^2+m^2} \, e^{i |\vec{q}| r}
\,\,\, .
\label{eq:oyuk}
\end{equation}
The standard approach is to evaluate the $|\vec{q}|$ integral via the residue theorem using a closed contour in the complex plane that encloses a pole at $|\vec{q}|=i \,m$, taking into account that the circular contour at infinity in the upper-half plane vanishes.  
However, in the present scenario, this latter integral is modified
\begin{equation}
V(\vec{x}) =-\frac{g^2}{4 \pi^2\, i r} \int_{-\infty}^\infty  d |\vec{q}| |\vec{q}|  \frac{
e^{-\eta |\vec{q}|^2}}{|\vec{q}|^2+m^2} \, e^{i |\vec{q}| r}  \,\,\, ,
\end{equation}
and the contour at infinity does not vanish everywhere due to the exponential factor. Hence, we must use a different approach.  We first exponentiate the denominator using a Schwinger parameter $u$,
\begin{equation}
V(\vec{x}) =-\frac{g^2}{4 \pi^2\, i r} \int_0^\infty du \, e^{-u\, m^2} \int_{-\infty}^\infty d |\vec{q}| |\vec{q}| 
\exp\{-(\eta+u) |\vec{q}|^2 + i |\vec{q}|  r \}\,\,\,. 
\end{equation}
The integral in $|\vec{q}|$ is of a recognizable form and can be done analytically:
\begin{equation}
V(r) = -\frac{g^2}{8 \pi^{3/2}} \, \int_0^\infty du \, \frac{1}{(\eta+u)^{3/2}}\, \exp\left\{-u \,m^2 - \frac{r^2}{4(\eta+u)}\right\} \,\,\,.
\label{eq:uint}
\end{equation}
The integral in Eq.~(\ref{eq:uint}) is probably not in a recognizable form for most, but nonetheless can also be done analytically.  The 
result is
\begin{equation}
V(r) = -\frac{g^2}{4\pi}\frac{1}{r} \,e^{\eta\, m^2} \left\{ - \sinh m r +
\frac{1}{2}\left(e^{m r}\, {\rm Erf}\left[\frac{r+2 m \eta}{2 \sqrt{\eta}}\right] + e^{- m r}\, {\rm Erf}\left[\frac{r-2 m \eta}{2 \sqrt{\eta}}\right] \right)\right\} \, .
\label{eq:nonlocpot}
\end{equation}
Note that for any finite $r$, the error functions become unity as $\eta \rightarrow 0$, so that the quantity in curly
brackets becomes $\cosh(m r) - \sinh(m r) = \exp(-m r)$, which yields
\begin{equation}
\lim_{\eta \rightarrow 0} V(r) \equiv V_0(r)=  -\frac{g^2}{4\pi}\frac{1}{r} e^{-m r} \,\,\, ,
\end{equation}
the usual result for a Yukawa potential.  The shape of the potential for various choices of $\eta$ is shown in 
Fig.~\ref{fig:one}.
This figure illustrates two qualitative features:  (1) the presence of the exponential eliminates the singularity at the origin; the potential
is regular at that point, approaching a constant up to corrections of ${\cal O}(r^2)$, and (2) the ``smoothing out" of the potential due to the 
nonlocality increases its depth for $r \agt m$ compared to the $\eta=0$ case.  Note that the quantity in curly brackets in Eq.~(\ref{eq:nonlocpot}) 
approaches $\exp(-m r)$ in the limit $\eta^{-1/2}\, r \rightarrow \infty$, for finite $\eta$.  In this case, the potential has the Yukawa form, with an extra 
multiplicative factor of $\exp(\eta \, m^2)$.  Of course, if measurement of the coupling $g$ happens only via this potential, then this factor 
could be absorbed in a redefinition of the coupling; however, if $g$ is measured in another process, then this difference in normalization 
might also be discernible.

Finally, we comment on the effects of motion relative to the preferred frame, defined by a velocity vector $\vec{v}$, which introduces a preferred direction.  The effect on our previous calculation is to take the exponentiated factor $\eta\, \delta_{ij} q^i q^j$ and replace it with
\begin{equation}
\eta \left[ \delta_{ij} +  \gamma^2 \,v_i v_j \right] q^i q^j \,\,\, ,
\end{equation}
where $\gamma$ is the usual relativistic factor $(1-v^2)^{-1/2}$.  Here we first have performed a Lorentz boost in 
the $\vec{v}$ direction, and have applied the usual nonrelativistic approximation (in the new frame) in which the $t$-channel 
momentum transfer has $q^0=0$, up to negligible corrections.  Since we have no knowledge {\em a priori} of the 
vector $\vec{v}$, phenomenological constraints on new terms in the potential generated by this boost can be interpreted as providing upper bounds 
on its magnitude $|\vec{v}|$.  However, as we alluded to earlier, if we were to assume that the preferred frame corresponds to one in which the cosmic 
microwave background is isotropic, then observations would tell us that $|\vec{v}| \approx 0.0012$ in units where $c=1$, corresponding to the measured 
value $369.82 \pm 0.11$ km/s~\cite{Tanabashi:2018oca}.  For small velocities like this, it is reasonable to calculate the effect on $V(r)$ given in 
Eq.~(\ref{eq:nonlocpot}) by expanding to quadratic order in $v$.   From our new starting point,
\begin{equation}
V(\vec{x},\vec{v}) = -g^2 \, \int \frac{d^3 q}{(2 \pi)^3} \frac{e^{-\eta |\vec{q}|^2- \eta \gamma^2 (v\cdot q)^2}}{|\vec{q}|^2+m^2} \, e^{i \vec{q}\cdot \vec{x}} \,\,\, ,
\label{eq:modified}
\end{equation}
we may write
\begin{equation}
V(\vec{x},\vec{v}) = \exp\left[ \eta \gamma^2 v^i v^j \partial_i \partial_j \right] \, V(\vec{x},\vec{0})
\end{equation}
which is useful only in that we are expanding the differential operator to second order in $\vec{v}$:
\begin{equation}
V(\vec{x},\vec{v}) = \left[1+ \eta \,v^i v^j \partial_i \partial_j \right] \, V(\vec{x},\vec{0}) + {\cal O}(v^4) \,\,\, .
\label{eq:v2}
\end{equation}
This allows us to find the desired $\vec{v}$ dependence using what we have already found in Eq.~(\ref{eq:nonlocpot}).  Since
$V(\vec{x},\vec{0}) \equiv V(r) $ depends only on $r$, we may rewrite Eq.~(\ref{eq:v2}) in terms of derivatives with respect to the radial coordinate:
\begin{equation}
V(r,\vec{v}) = V(r) +\eta \, |\vec{v}|^2 \frac{1}{r} \frac{dV(r)}{dr} -  \eta\,  \frac{(\vec{v}\cdot\vec{x})^2}{r^2} \left( \frac{1}{r} \frac{dV(r)}{dr} - \frac{d^2 V(r)}{d r^2} \right) + {\cal O}(v^4) \,\,\, .
\end{equation}
The first correction term, reading from left to right above, is spherically symmetric and non-singular at the origin; it simply represents a small correction to the radial 
potential that we have already discussed, suppressed by at least a factor of $|\vec{v}|^2 \sim 10^{-6}$ if the CMB defines the preferred frame.  The second correction term
is qualitatively different, since it is sensitive to the preferred direction.   Possible corrections to the gravitational potential proportional to $(\vec{v}\cdot\vec{x})^2/r^3$ (which
has a different radial dependence) must be suppressed below gravitational strength by a factor $\alpha_2 = 2 \times 10^{-9}$, where $\alpha_2$ is defined in the 
Parameterized Post Newtonian, or PPN, formalism~\cite{Will:2014kxa}.  However, this bound,  which is determined from the precession of pulsar rotation axes, does not 
apply here since it assumes a force with infinite range.   We will assume henceforth that the range of our new force is substantially less than $10^4$~meters 
($m \gg 10^{-11}$~eV), the size of a typical neutron star, so that an analogous bound is evaded.   Whether interesting astrophysical bounds on Lorentz-violating forces 
with finite range can be determined is worthy of investigation, but will not be considered in the present work.

\section{Nonlocal Lorentz-violation in a dark sector} \label{sec:darksector}
In this section, we consider how a single real scalar field, like the one discussed in the previous section, might couple to 
matter fields of the standard model in a realistic scenario.  We do not identify the scalar field as dark matter, but 
assume that it could decay into other dark-sector particles that are stable or suitably long lived.  The portal to the standard 
model will consist of a sector of heavy vector-like fields.   Let us first discuss the portal in the case of a local theory and 
then explain how we introduce the nonlocality into the theory in a way that will keep the Lorentz-violation suitably sequestered.

Consider a heavy, vector-like field ${\cal D}$ with the same quantum numbers as a right-handed down quark $d_R$:
\begin{equation}
{\cal D}_R \sim {\cal D}_L \sim d_R  \,\,\, .
\end{equation}
The mass terms and Yukawa couplings that involve these fields are the following:
\begin{equation}
{\cal L} = - M \overline{{\cal D}}_L \, {\cal D}_R - \overline{Q}_L \, H \, d_R - \overline{Q}_L\, H \, {\cal D}_R
-\phi \, \overline{{\cal D}_L} \, {\cal D}_R -\phi \, \overline{{\cal D}_L} \, d_R + \mbox{H.c.} 
\label{eq:portal}
\end{equation}
Here we have suppressed the dimensionless couplings and considered standard model quarks $Q_L$ and $d_R$ 
of a single generation, for simplicity.   Note also that a mixing term of the form $\Delta m\, \overline{{\cal D}_L} \, d_R$ has been
eliminated by a definition of the ${\cal D}_R$-$d_R$ field basis.
\begin{figure}[t]
\begin{center}
    \includegraphics[width=.5\textwidth]{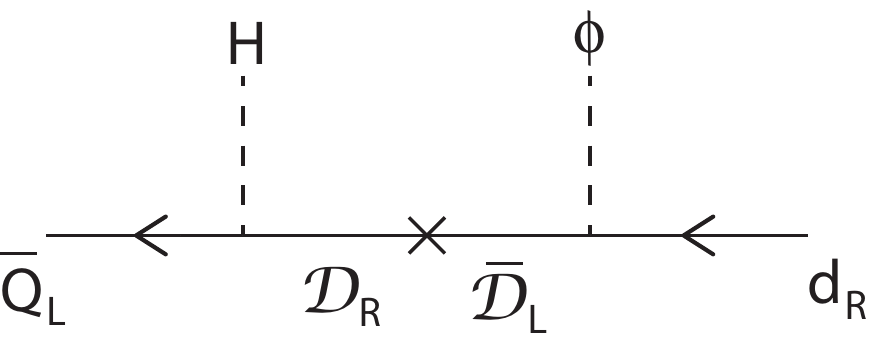}
    \caption{Diagram involving the exchange of the vector-like quark ${\cal D}$.}
        \label{fig:two}
  \end{center}
 \end{figure}
When the heavy ${\cal D}$ fields are integrated out of the theory, higher-dimension operators will be generated in the low-energy
effective theory.   In the lingo of Froggatt-Nielsen model building~\cite{Froggatt:1978nt}, one operator of interest is generated via the ``spaghetti" diagram 
shown in Fig.~\ref{fig:two}.   The amplitude for this diagram in momentum space
\begin{equation}
i {\cal M} =  \overline{u}_Q \left[ (i P_R) \frac{i\, (\slashed{p}+M)}{p^2-M^2} (i P_R) \,\right] u_d \longrightarrow
 i \frac{1}{M}\, \overline{u}_Q \, P_R \, u_d \,\,\, ,
\label{eq:spamp}
\end{equation}
where $p$ is the momentum on the internal line, $P_R = (1+\gamma^5)/2$.  On the far right of Eq.~(\ref{eq:spamp}) we show the limit
in which $p^2 \ll M^2$.   In the low-energy effective theory, this amplitude is reproduced by the higher-dimension operator
\begin{equation}
{\cal L}_{eff} = \frac{\kappa_0}{M} \phi \,  \overline{Q_L} \, H \, d_R \,\,\, ,
\end{equation}
where $\kappa_0$ subsumes the product of all the undetermined Yukawa couplings relevant to the diagram of Fig.~\ref{fig:two}.  This leads to
a Yukawa interaction of the same form that we assumed in Sec~\ref{sec:potential}, with coupling $\kappa_0 \, v / (\sqrt{2} M)$, where $v=246$~GeV is 
the Higgs vev.

We now introduce the desired nonlocality by modifying only the two terms in Eq.~(\ref{eq:portal}) that depend on $\phi$:
\begin{equation}
-\overline{{\cal D}_L} \, {\cal D}_R \, e^{\eta \nabla^2/2} \phi - \overline{{\cal D}_L} \, d_R \, e^{\eta \nabla^2/2} \phi + \mbox{H.c.}
\label{eq:modform} 
\end{equation}
Notice that a field redefinition $\phi = e^{- \eta \nabla^2/2} \varphi$ would move this nonlocal factor back to the quadratic terms of $\varphi$, as in the toy 
models we have considered earlier, as well as to other possible interaction terms.  The value of introducing the nonlocality initially in Eq.~(\ref{eq:modform}) is 
that it makes clear that the Lorentz-violation involves the super-heavy field ${\cal D}$; after integrating out the heavy sector, Lorentz-violating effects in the low-energy
effective theory will always be suppressed by a ratio of widely separated mass scales.   In other words, our assumption of adequate sequestering dictates where 
we introduce the factors of $\hat{F}$ in the Lagrangian.  Had we instead introduced $\hat{F}^{-1}$ initially in the quadratic terms, and left 
Eq.~(\ref{eq:portal}) unchanged, then one might have Higgs sector Lorentz violation suppressed only by loops involving the possible renormalizable couplings 
between $\phi$ and $H$.  One might formulate the theory in that way if there is a reason to expect those Higgs portal couplings to be absent, for example, 
if  $\phi$ and $H$ are separated in an extra dimension, while ${\cal D}$ is a bulk field.  Such an extra dimensional formulation may be desirable since it would 
also eliminate Planck-suppressed higher-dimension operators that couple the $\phi$ field directly to standard model fields, for example 
$\phi F_{\mu\nu} F^{\mu\nu}/M_P$, which leads to additional constraints~\cite{Calmet:2020rpx}.  For the purpose of our present discussion, we assume that such 
operators, if present, are adequately suppressed. 

The construction just described can be extended by introducing another heavy, vector-like field ${\cal U}$ with the same quantum numbers as a right-handed up-quark.  We would then generate the higher-dimensional operators
\begin{equation}
{\cal L}_{eff} = \frac{\kappa^d_0}{M} e^{\eta \nabla^2/2} \phi \,  \overline{Q_L} \, H \, d_R + \frac{\kappa^u_0}{M} e^{\eta \nabla^2/2} \phi \,  \overline{Q_L} \, \widetilde{H} \, u_R\,\,\, ,
\label{eq:hdop}
\end{equation}
and, after setting $H$ to its vev, the induced Yukawa couplings
\begin{equation}
{\cal L}_{yuk} = \sum_{f=u,d} \frac{\kappa_f \, m_f}{M} \,\overline{f} \, f \, e^{\eta \nabla^2/2} \phi \,\,\,.
\label{eq:ffp}
\end{equation}
Here $\kappa_f \equiv \kappa_0^f/\lambda_f$, where $\lambda_f$ is the standard model Yukawa coupling  $\sqrt{2} m_f / v$.   Thus, we have 
defined  $\kappa_f$ to be unity if it is of the same size as the dimensionless coupling we would associate with either $\overline{Q}_L \, H \, d_R$ or 
$\overline{Q}_L \, \widetilde{H} \, u_R$, operators with a similar flavor structure.   This provides a convenient point of reference.

The location of the exponential factor in Eq.~(\ref{eq:ffp}), which appears when $\phi$ has canonical kinetic terms, yields the same nonrelativistic 
potential as the one considered in Sec.~\ref{sec:potential}.  With Eq.~(\ref{eq:ffp}) at hand, another useful effective interaction to consider is the coupling 
of $\phi$ to nucleons, 
\begin{equation}
{\cal L}_{eff} = f_p \, \overline{p}\, p \, e^{\eta \nabla^2/2}\phi  + f_n  \,\overline{n}\, n\, e^{\eta \nabla^2/2}\phi \,\,\, .
\label{eq:nuccoup}
\end{equation}
The mapping from Eq.~(\ref{eq:ffp}) to Eq.~(\ref{eq:nuccoup}) is the same as found in studies of scalar dark matter.  From
Ref.~\cite{Ellis:2000ds}, 
\begin{equation}
\frac{f_N}{m_N} = \sum_{u,d,s} f_{T_q}^{(N)} \frac{\alpha_q}{m_q}  + \frac{2}{27} \, f_{TG}^{(N)} \sum_{c,b,t} \frac{\alpha_q}{m_q}  \,\,\,,
\end{equation}
where the scalar-quark couplings in this case are given by
\begin{equation}
\alpha_q = \left\{ \begin{array}{cl} \frac{\kappa_q \, m_q}{M} & \mbox{for } q=u,d \\ 0 & \mbox{otherwise} \end{array}\right.
\label{eq:alphas}
\end{equation}
Note that Eq.~(\ref{eq:alphas}) reflects the fact that the simple model we have presented only provides for couplings to the lightest two quark flavors;  
however, it is straightforward to extend the vector-like sector so that couplings of $\phi$ to heavier flavors are induced as well.  Numerical values 
of $f_{T_q}^{(N)}$ and $f_{TG}^{(N)}$, for $N=p$ or $n$, can be found in Ref.~\cite{Ellis:2000ds}.  For the purpose of an estimate, we will further
assume that $\kappa_u = \kappa_d \equiv \kappa$.   We find
\begin{equation}
{\cal L}_{eff} = {\cal A}_p   \kappa \frac{m_p}{M} \, [\overline{p} \, p \,  e^{\eta \nabla^2/2}\phi]
+ {\cal A}_n   \kappa \frac{m_n}{M} \, [\overline{n} \, n \,  e^{\eta \nabla^2/2}\phi]  \,\,\,,
\label{eq:ncoup}
\end{equation}
where ${\cal A}_p = 0.046$ and ${\cal A}_n=0.050$.  We then infer that the potential due to $\phi$-exchange between two atoms
with atomic number $Z$ and atomic mass $A$ is given by Eq.~(\ref{eq:nonlocpot}) with the replacement 
\begin{equation}
\frac{g^2}{4 \pi} \rightarrow \frac{\kappa^2}{4 \pi \, M^2} \left[Z ({\cal A}_p m_p - {\cal A}_n m_n) + {\cal A}_n m_n A \right]^2  \,\,\, .
\end{equation}
Note that the factor ${\cal A}_p m_p - {\cal A}_n m_n$ would vanish in the absence of isospin breaking effects and is suppressed
relative to the second term in brackets.   For example, for iron, where $Z=26$ and $A=56$, the first term represents a $3.8\%$ effect.
For the purposes of an estimate, we ignore isospin differences, so that the potential between two atoms is given by 
Sec.~\ref{sec:potential} as follows
\begin{equation}
V(r) = -\frac{\kappa^2{\cal A}_N^2}{4 \pi \, M^2}   \frac{M_a^2}{r} \,e^{\eta\, m^2} \left\{ - \sinh m r +
\frac{1}{2}\left(e^{m r}\, {\rm Erf}\left[\frac{r+2 m \eta}{2 \sqrt{\eta}}\right] + e^{- m r}\, {\rm Erf}\left[\frac{r-2 m \eta}{2 \sqrt{\eta}}\right] \right)\right\} \, .
\label{eq:actualv}
\end{equation}
where $M_a$ is the mass of each atom, and ${\cal A}_N \approx 0.05$.   As we have discussed earlier, this potential becomes Yukawa-like asymptotically,
so we can obtain an estimate of the typical bounds on $\kappa$ using the results in Ref.~\cite{Leefer:2016xfu}, which apply to a Yukawa-like force.  The scale 
suppression $1/\Lambda$ in this reference can be matched to ${\cal A}_N \kappa / M$ in Eq.~(\ref{eq:ncoup}).   If we take, for example, $M=0.1\, M_P$,
we find $\kappa < 0.02$, for a force with a range below $10^4$~m (or $m > 10^{-11}$~eV).  (This would become  $\kappa < 0.01$ if one corrects $\kappa^2$ by the 
nonlocal factor $e^{\eta m^2}$, for $\eta m^2=1$.)   If we define a parameter $\xi$  that compares the coefficient of Eq.~(\ref{eq:actualv}) to 
gravitational  strength, {\i.e.}, $\kappa^2{\cal A}_N^2 e^{\eta m^2}/(4 \pi \, M^2) = \xi /M_P^2$, where $M_P=1.2 \times 10^{19}$~GeV is the Planck mass, then for this choice
of parameters, with $\kappa < 0.01$,  one finds $\xi< 5 \times 10^{-6}$.  This is consistent with the statement in Ref.~\cite{Will:2014kxa} that bounds range 
from $10^{-3}$ to $10^{-6}$ the strength of gravity for ranges between $1$ and $10^4$~m.   In any case, we will assume that the upper bounds on 
$\kappa$ are satisfied so that our theory remains consistent with fifth force searches.   We note that astrophysical bounds on very light scalars are superseded by fifth force bounds for scalar masses below $0.2$~eV~\cite{Hardy:2016kme}, and will not provide additional constraints.
  
Separate bounds come from the fact that the theory is Lorentz-violating.  For example, the interaction in Eq.~(\ref{eq:ffp}) provides a 
Lorentz-violating contribution to the self-energy function for the fermion $f=u$ or $d$, which leads to a Lorentz-violating dispersion relation.  
We can use this to compute the difference between the speed of a massless fermion and the speed of light, a quantity often used to constrain theories with Lorentz-violation that is isotropic~\cite{Pospelov:2010mp}. From Eq.~(\ref{eq:ffp}), the self-energy (following the conventions of Peskin and Schroeder~\cite{Peskin:1995ev}) is given by
\begin{equation}
-i \Sigma = g_f^2  \int_0^1 dx \, \int \frac{d^4 q}{(2\pi)^4} 
\frac{e^{-\eta(\vec{q}-x \, \vec{p})^2}\left[\slashed{q}+(1-x)\slashed{p}+m_f\right]}{[q^2- \Delta]^2} \,\,\, ,
\label{eq:sig}
\end{equation}
where $g_f = \kappa_f \, m_f / M$ and
\begin{equation}
\Delta = -x\, (1-x) \, p^2 + (1-x) \, m^2 +x \, m_f^2 \,\,\, .
\end{equation}
We show we can obtain  a useful bound by studying the limit in which the dimensionless quantity $\eta^{1/2} \vec{p}$ is small, and 
looking at the corrections to the fermion dispersion relation that are obtained at first order in this quantity.   In the Appendix, we consider 
the more general case, and confirm that the final result of this section can be obtained without using an expansion.  Expanding the integral in Eq.~(\ref{eq:sig}), the self-energy function takes the form
\begin{equation}
\Sigma = -{\cal A} \, \slashed{p} + {\cal B} \, m_f + {\cal C} \, \vec{p} \cdot \vec{\gamma} + \cdots \,\,\ ,
\end{equation}
where the $\cdots$ refers to terms suppressed by an additional power of $\eta^{1/2} \vec{p}$, and where ${\cal A}$, ${\cal B}$ and
${\cal C}$ are the following dimensionless, Euclidean integrals
\begin{equation}
{\cal A} = g_f^2 \int_0^1 dx \, (1-x) \int \frac{d^4 q_E}{(2\pi)^4} \frac{e^{-\eta |\vec{q}\,|^2}}{(q_E^2 + \Delta)^2} \,\,\, ,
\end{equation}
\begin{equation}
{\cal B} = - g_f^2 \int_0^1 dx \int \frac{d^4 q_E}{(2\pi)^4} \frac{e^{-\eta |\vec{q}\,|^2}}{(q_E^2 + \Delta)^2} \,\,\, ,
\end{equation}
\begin{equation}
{\cal C} = \frac{2}{3}\, g_f^2 \int_0^1 dx \, x \int \frac{d^4 q_E}{(2\pi)^4} \frac{\eta |\vec{q}\,|^2\,e^{-\eta |\vec{q}\,|^2}}{(q_E^2 + \Delta)^2} \,\,\, ,
\label{eq:cc}
\end{equation}
with $q_E^2=(q^0)^2+|\vec{q}\,|^2$.   The condition for an on-shell fermion
\begin{equation}
\slashed{p}-m_f - \Sigma(\slashed{p})=0 
\end{equation}
can thus be written as
\begin{equation}
(1+{\cal A})\, \slashed{p} - (1+{\cal B}) \, m_f - {\cal C} \, \vec{p} \cdot \vec{\gamma} =0  \,\,\, .
\end{equation}
Multiplying both sides of this expression by quantity that is the same as the left-hand-side with the sign of the second term flipped 
allows us to eliminate the gamma matrix structure,
\begin{equation}
(1+2 {\cal A})\, p^2 - (1+2 {\cal B})\, m_f^2 - 2 {\cal C} \, |\vec{p}\,|^2=0 \,\,\, ,
\end{equation}
where we have dropped negligible terms that are of order $g_f^4$.  We may solve for $p^0$, perturbatively in $g_f^2$, to
obtain the dispersion relation
\begin{equation}
(p^0)^2 = (1+2 \,{\cal \tilde{C}})\, |\vec{p}\,|^2 + (1-2 \, {\cal \tilde{A}}+2\, {\cal \tilde{B}})\, m_f^2 \,\,\, ,
\end{equation}
where the tilde indicates our previous expressions with the function $\Delta(p^2)$ evaluated at $p^2=m_f^2$. In the limit that the fermion is 
massless, its speed $c_0$  can be read off the first term 
\begin{equation}
c_0 = 1 + {\cal \tilde{C}}  \,\,\, ,
\end{equation}
again working to order $g_f^2$, and where we have set the speed of photons $c=1$.   The quantity $|c - c_0|$ is
experimentally bounded~\cite{Lamoreaux:1986xz}, such that
\begin{equation}
{\cal \tilde{C}} <  3 \times 10^{-22} \,\,\, ,
\end{equation}
where we have used the fact that ${\cal \tilde{C}}>0$.  More explicitly, this can be written
\begin{eqnarray}
&& \frac{g_f^2}{12 \pi^2} \left[ \int_0^1 dx \, x \int_0^\infty dy \, \frac{y^4 e^{-y^2}}{[y^2+ (1-x) \,  \rho ]^{3/2}} \right]  \nonumber \\
&& =\frac{g_f^2}{12 \pi^2} \left[\frac{1}{2 \rho^2} \left\{8 + 2 \rho - 3 \sqrt{\pi} \, U(-\frac{1}{2},-2,\rho) \right\}\right] < 3 \times 10^{-22} \,\,\, ,
\label{eq:intbnd}
\end{eqnarray}
where $\rho \equiv \eta \, m^2$ and $U(a,b,z)$ is the confluent hypergeometric function~\cite{confluent}.  The integral in the first line of Eq.~(\ref{eq:intbnd}) can be obtained from Eq.~(\ref{eq:cc}) by performing the $q^0$ and $\vec{q}$ angular integrations, so that the 
Feynman parameter integral and a radial $|\vec{q}|$ integral remain.  The quantity in square brackets never exceeds $1/4$ for any 
nonnegative $\rho$; from this, we obtain the bound $g_f = \kappa \, m_f / M < 3.8 \times 10^{-10}$.  Had we done this calculation using 
the effective interaction for the proton that we derived earlier, $\kappa m_f$ would be replaced by $\kappa {\cal A}_N m_p$.   In either 
case, the ratio of mass scales (for example $1$~GeV$/ [0.1 M_P] \sim 10^{-18}$) by itself assures that the bound is satisfied and is 
superseded by the bounds that we discussed earlier on long range forces.  

We note in the Appendix that away from the limit considered in this section, ${\cal \tilde{C}}$ is a function of $3$-momentum that drops off 
quickly with increasing $|\vec{p}|^2$, so that its effects become suppressed.  As the case of small $\eta^{1/2} \vec{p}$ provides
Lorentz-violating effects that are maximal but does not give additional constraints on our theory, we will not study the unusual form of the 
dispersion relation for arbitrary momenta here.  That issue, as well as a more general study of Lorentz-violating effects in similar nonlocal 
theories will be considered in a separate publication.
 
\section{Conclusions} \label{sec:conc}
In this paper, we have considered scalar theories in which quadratic terms are present that are nonlocal and Lorentz-violating.  Part of our initial
motivation was to avoid the complications related to unitarity discussed in the related Lorentz-invariant theories of Ref.~\cite{Carone:2016eyp}; we 
verified this by repeating the same calculation presented in that earlier work.  However, the theories discussed here are potentially of interest for a 
broader set of reasons.  For example, they suggest a nonlocal generalization of the Horava-Lifshitz  idea~\cite{Horava:2009uw}, and might 
be useful in formulating a renormalizable quantum theory of gravity.   Moreover, as indicated qualitatively by the smoothing of singularities at the 
origin of the nonrelativistic potential studied in Sec.~\ref{sec:potential}, the nonlocality we discussed might capture some features of an underlying 
ultraviolet completion.   

Since the violation of Lorentz invariance must confront stringent experimental bounds~\cite{Tasson:2014dfa}, the scale of nonlocality can only be low
in sectors that communicate very weakly with standard model particles.  We have focused on that possibility here, assuming by necessity that any nonlocal 
modification of the standard model Lagrangian itself occurs at much shorter distance scales.  While gravity provides one possible avenue for exploration, in 
the present work we considered the possibility of a ``dark" scalar sector that couples to the standard model through an ``portal" sector of heavy, vector-like 
fermions.   When the heavy fermions are integrated out of the theory, the couplings induced to standard model fermions are extremely weak.  Nevertheless, 
ultralight particles from a nonlocal, Lorentz-violating sector may be detectable via the long-range forces that they mediate, which could lead to detectable 
corrections to the gravitational potential of macroscopic bodies.  The sequestering of this sector allows the scale of nonlocality to be comparable to the 
mass scale of the particle mediating the long-range force, a possibility that has not been considered previously in the literature.

In summary, this paper has proposed a new possibility, that of nonlocal Lorentz-violating extensions of the standard model.   Although we have constructed 
one explicit model and considered some aspects of its phenomenology, the more important, overarching point is that the general idea presented here 
may lead to other interesting applications. Directions for future study could include a more general study of the Lorentz-violating effects in similar 
theories, and the development of a non-local Lorentz-violating modification of gravity.  A systematic study of the renormalization of Lorentz-violating 
nonlocal theories would also be worthwhile.  We hope to return to these topics in future work.

\begin{acknowledgments}  
We thank the NSF for support under Grant PHY-1819575.  We are grateful to Josh Erlich and Jens Boos for their valuable 
comments.  
\end{acknowledgments}

\appendix
\section{} \label{sec:appendix}

In this Appendix, we briefly outline the evaluation of Eq.~(\ref{eq:sig}), without expanding in $\eta^{1/2} \vec{p}$.  First, we note that
we may evaluate $\Sigma$ at the point $p^2=m_f^2$, since corrections to the dispersion relation affect $\Sigma$ at higher-order
in $g_f^2$.  In this case, the factor of $q_E^2 + \Delta$ that appears in the Euclideanized denominator of Eq.~(\ref{eq:sig}) is always 
positive, as $\Delta = x^2\, m_f^2 + (1-x)\, m^2 > 0$.  We are therefore justified in exponentiating the denominator using  a Schwinger 
parameter $u$. We can then shift integration variables so that the quantity that is exponentiated in the integrand is spherically symmetric, 
which allows us to discard odd terms in $q$.  The momentum integrals are then all gaussian and can be easily evaluated.  When the dust settles, we are left with
\begin{equation}
\Sigma = -\frac{g_f^2}{16 \pi^2} \int_0^1 dx \int_0^\infty du \, \frac{u^{1/2}}{(\eta+u)^{3/2}}
\, e^{-\frac{\eta \, u \, x^2}{u+\eta} |\vec{p}|^2 - u \Delta} \left[-\frac{\eta\, x}{\eta+u} \, \vec{p}\cdot \vec{\gamma} + (1-x) \, \slashed{p}
+m_f\right] \,\,\, .
\label{eq:sigagain}
\end{equation}
Note that setting $\vec{p}$ to zero only in the exponential factor in Eq.~(\ref{eq:sigagain}) provides an upper bound for the value of the integrals, since the exponential is always less than $1$ over the integration region.  Doing so, and setting $m_f$ to zero,  we would identify
\begin{equation}
{\cal \tilde{C}} = \frac{g_f^2}{12 \pi^2} \left[\frac{3}{4} \int_0^1 dx \,  x \int_0^\infty d\tilde{u} \, \frac{\tilde{u}^{1/2}}{(1+\tilde{u})^{5/2}} e^{-(1-x) 
\tilde{u} \, \rho}  \right]\,\,\, ,
\label{eq:secform}
\end{equation}
where $\tilde{u} = \eta^{-1} u$ and $\rho$ is defined as in Sec.~\ref{sec:darksector}.  We have confirmed numerically that 
Eq.~(\ref{eq:secform}) and  Eq.~(\ref{eq:intbnd}) are identical.   It is also clear from Eq.~(\ref{eq:sigagain}) (and we have checked numerically) that the coefficient of the $\vec{p}\cdot \vec{\gamma}$ terms is a rapidly decreasing function of $|\vec{p}|^2$, as noted in 
Sec.~\ref{sec:darksector}.

\end{document}